# 2

# The Role of Head-Up Display in Computer-Assisted Instruction

Kikuo Asai
*National Institute of Multimedia Education*
*Japan*

## 1. Introduction

The head-up display (HUD) creates a new form of presenting information by enabling a user to simultaneously view a real scene and superimposed information without large movements of the head or eye scans (Newman, 1995; Weintraub and Ensing, 1992). HUDs have been used for various applications such as flight manipulation, vehicle driving, machine maintenance, and sports, so that the users improve situational comprehension with the real-time information. Recent downsizing of the display devices will expand the HUD utilization into more new areas.

The head-mounted display (HMD) has been used as a head-mounted type of HUDs for wearable computing (Mann, 1997) that gives a user situational information by wearing a portable computer like clothes, a bag, and a wristwatch. A computer has come to interact intelligently with people based on the context of the situation with sensing and wireless communication systems.

One promising application is in computer-assisted instruction (CAI) (Feiner, et al., 1993) that supports the works such as equipment operation, product assembly, and machine maintenance. These works have witnessed the introduction of increasingly complex platforms and sophisticated procedures, and have required the instructional support. HUD-based CAI applications are characterized by real-time presentation of instructional information related to what a user is looking at. It is commonly thought that HUD-based CAI will increase productivity in instruction tasks and reduce errors by properly presenting task information based on a user's viewpoint.

However, there are not enough empirical studies that show which factors of HUDs improve user performance. A considerable amount of systematic research must be carried out in order for HUD-based CAI to fulfill its potential to use the scene augmentation to improve human-computer interaction.

We have developed a HUD-based CAI system that enables non-technical staff to operate the transportable earth station (Asai, et al., 2006). Although we observed that users of the HUD-based CAI system performed slightly better than users of conventional PCs and paper manuals, it was not clear which factors significantly affected performance in operating the system. We here conducted a laboratory experiment in which participants performed a task of reading articles and answering questions, in order to evaluate how readable the display of the HUD is, how easy it is to search information using the system, and how it affects the work efficiency. We then discuss the characteristics of HUD-based CAI, comparing the task performance between the HUD-based CAI and conventional media.

Thus, this chapter is a study on the information processing behavior at an HUD, focusing on its role in CAI. Our aim is to introduce the basic concept and human factors of an HUD, explain the features of HUD-based CAI, and show user performance with our HUD-based CAI system.

## 2. HUD Technology

The HUD basically has an optical mechanism that superimposes synthetic information on a user's field of view. Although the HUD is designed to allow a user to concurrently view a real scene and superimposed information, its type depends on the application. We here categorize HUDs into three design types: head-mounted or ground-referenced, optical see-through or video see-through, and single-sided or two-sided types.

### 2.1 Head-mounted and Ground-referenced Types

HUDs are categorized into the head-mounted and ground-referenced types in terms of spatial relationship between the head and HUD, as shown in Fig. 1.

In the head-mounted type (Fig. 1 (a)), an HUD is mounted on the head, being attached to a helmet or a head band. It is generally called a head-mounted display (HMD). Since the HUD is fixated to the head, a user can see visual information, even though moving the head. The head-mounted type of HUD is used at the environment where users have to look around them, such as building construction, surgical operation, and sports activities. The head-mounted HUD should be light in weight, because the user has to support its weight.

In the ground-referenced type (Fig. 1 (b)), an HUD is grounded to a desktop, wall, or floor. Since the relation between the head and HUD is not fixated spatially, visual information can be viewed just in case that a user directs the head to the HUD. The ground-referenced type of HUD is used at the environment where users almost look at the same direction, such as flight manipulation and vehicle driving. The user does not need to support the weight in the ground-referenced HUD.

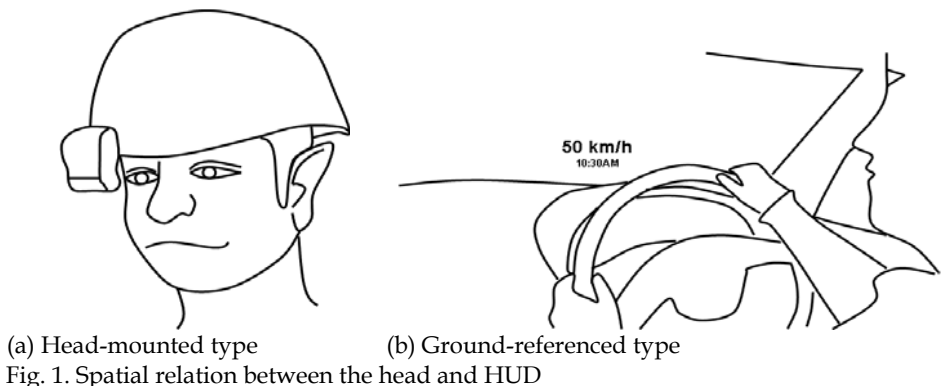

(a) Head-mounted type	(b) Ground-referenced type
Fig. 1. Spatial relation between the head and HUD

## 2.2 Optical See-through and Video See-through Types

HUDs are categorized into the optical see-through and video see-through types in terms of optical mechanism, as shown in Fig. 2.

In the optical see-through type (Fig. 2 (a)), a user sees the real scene through a half mirror (transparent display) on which the synthetic images including graphics or text are overlaid. The optical see-through HUD has advantages of seeing the real scene without degradation of the resolution and delay of the presentation. In addition, eye accommodation and convergence responses work for the real scene. However, the responses do not work for virtual objects. That is, the real scene and the synthetic images are at different distances from the user. Therefore, the user's eyes need to alternately adjust to these distances in order to perceive information in the both contexts. Frequent gaze shifting to different depths may result in eye strain (Neveu, et al., 1998). The optical see-through HUD does not also represent occlusion correctly because the real scene goes through the half mirror at the pixel area of the front virtual objects. One more problem is of difficulty in use under a bright illumination condition such as an outdoor field because of low luminance of the display.

In the video see-through type (Fig. 2 (b)), a real scene is captured by a camera. The user sees the real scene images, in which information such as graphics or text is superimposed, at a display monitor. The video see-through HUD has the following advantages; (1) the real scene can be flexibly processed at the pixel unit, making brightness control and color correction, (2) there is no temporal deviation between the real scene and virtual objects because of their synchronous presentation in the display image, and (3) the additional information is obtained by using the captured scene, deriving depth information from parallax images and user's position from the geometric features. According to (1), the occlusion is achieved by covering the real scene with the virtual objects or culling the back pixels out of the virtual objects. While, the video see-through HUD has shortcomings due to losing rich information on the real scene. Low temporal and spatial resolution of the HUD decreases the realistic and immersive sense of the real scene. The inconsistent focus-depth information may result in high physiological load during the use. Despite (2), the video see-through HUD has presentation delay due to the image composition and rendering, which may sometimes lead to a critical accident at the environment such as a construction site.

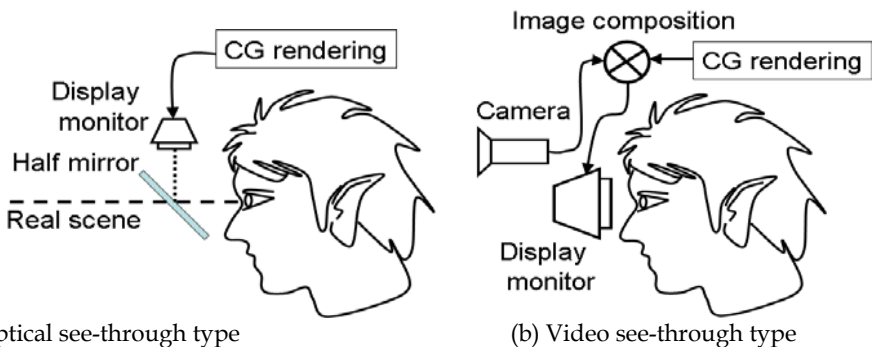

(a) Optical see-through type                    (b) Video see-through type
Fig. 2. Optical difference between the HUDs

## 2.3 Single-sided and Two-sided Types

HUDs are categorized into the single-sided and two-sided types in terms of the field of view based on relation of the eyes and HUD, as shown in Fig. 3. Whether presenting the synthetic images to one eye or two eyes is an important factor that dominates the range of applicable areas.

In the single-sided type of HUD (Fig. 3 (a)), the real scene is viewed by two eyes, and the synthetic images are presented to one eye using an optical see-through or small video see-through display. The real scene images captured by a video camera have a time lag to be displayed at the video see-through display. A single-sided HUD is used at the environment where a user works looking at the peripheral situations or experience of the real world proceeds acquisition of the complementary information. For example, the single-sided type is usually used in a construction site due to safety reasons. When the synthetic images or the device frames interfere largely with the user's field of view, vital accidents may occur during the work.

In the two-sided type of HUD (Fig. 3 (b)), the real scene and synthetic images are viewed by two eyes using an optical see-through or video see-through display. A two-sided HUD is used at the situation where safety of the user is ensured without looking around, because the cost of visual interference would be high at the two-sided HMD, in which the overlaid information interferes with the view of the workspace. For example, the two-sided type is often used at an entertainment situation because of producing the feeling of being there.

There is a tradeoff relationship between the single-sided and two-sided types in readability of documents on the HUD and visibility of the real scene via the HUD. The single-sided HUD enables a user to easily see real objects using one eye with no occlusion, though the user has to read documents using only one eye. On the other hand, the two-sided HUD enables the user to read documents with both eyes, though the user has to view real objects through the display on which the documents are presented. The single-sided HUD is more convenient for acquiring information on real objects, and the two-sided HUD is more convenient for acquiring information on the display.

In the head-mounted type, the weight of the HUD is an important factor for user's comfort. A single-sided HUD, in general, weighs less than the two-sided HUD. Although the difference in weight is only 150 g for the HUDs, it turns out to be significant because the device is attached to the head (Asai, et al., 2005). The heavier the HUD is, the tighter the HUD has to be placed on the head without being shifted, which may result in difficulty for a long-time use.

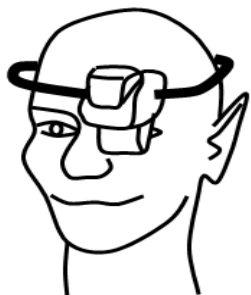    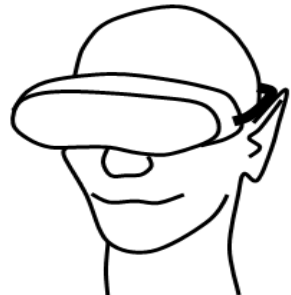

        (a) Single-sided type        (b) Two-sided type
Fig. 3. Relation between the eyes and HUD

## 3. Human Factors of HUDs

HUD systems have developed for improving performance of multiple tasks in aircraft landing and vehicular driving. In the aircraft landing, the HUD system supports pilots to keep operation performance in navigating through a crowded airspace. In the vehicular driving, the HUD supports drivers to keep driving performance in accessing information from multiple sources such as speed, navigation, and accidents. Although numerous information and communication tools have provided a user with a large amount of information, the user's capacity to process information does not change.

There are many researches regarding the costs and benefits of HUDs compared with head-down displays (HDDs). The benefits of HUDs are mainly characterized by visual scanning and re-accommodation. In the visual scanning, HUDs reduces the amount of eye scans and head movements required to monitor information and view the outside world (Haines, et al., 1980; Tufano, 1997). The traditional HDD causes time sharing between the tasks. For example, drivers must take their eyes off the road ahead in order to read the status at the control panel, which affects driving safety. The HUD degrades the problem because of simultaneous viewing of the monitor information and real scene. In the visual re-accommodation, HUDs reduces the adjustment time of refocusing the eyes required to monitor information and view the outside world (Larry and Elworth, 1972; Okabayashi, et al., 1989). The HDD makes the user refocus the eyes frequently for viewing the closer and far domains, which may cause fatigue. The HUD degrades the re-accommodation problem by allowing the user to read the status without shifting focus largely in case being optically focused farther.

However, use of HUDs did not always improve user performance in aviation safety studies, especially when unexpected events occurred (Fischer, et al., 1980; Weintraub, et al., 1985). The HUD users had a shorter response time than the HDD users to detect unexpected events only in conditions of low workload. The benefits of HUDs, however, were reduced or even reversed in conditions of high workload (Larish and Wickens, 1991; Wickens, et al., 1993; Fadden, et al., 2001). Measurement of the time required to mentally switch between the superimposed HUD symbology and the moving real-world scene revealed that it took longer to switch when there was differential motion between superimposed symbology in a fixed place on the HUD and movement in the real-world scene behind the symbology caused by motion of the aircraft (McCann, et al., 1993). As a result, conformal imagery (Wickens and Long, 1995) or scene-linked symbology (Foyle, et al., 1995) that moved as the real objects moved was configured on the HUD to reduce the time it takes to switch attention. The HUD can depict virtual information reconciled with physically viewable parts in the real world. This conformal imagery or scene-linked symbology is based on the same concept as spatial registration of virtual and real objects in augmented reality (AR) technology (e.g., Milgram and Kishino, 1994; Azuma, 1997).

## 4. HUD-based CAI

CAI has been applied to maintenance and manufacturing instructions in engineering (Dede, 1986), in which complex tasks must be performed. CAI systems have helped new users learn how to use devices by illustrating a range of functional capabilities of the device with multimedia content. However, the human-computer interfaces were inadequate for eliciting the potential of human performance, due to the limitations of the input/output devices, including inconvenience of holding a CAI system while working and mismatch of information processing between computer and human. An HUD environment may make it possible to improve human-computer interaction in CAI by allowing information to be integrated into the real scene.

## 4.1 Applications

Typical examples of HUD-based CAI applications are operation of equipment, assembly of products, and maintenance of machines. Many systems have been developed as applications of AR technology, including assistance and training on new systems, assembly of complex systems, and service work on plants and systems in industrial context (Friedrich, 2002; Schwald and Laval, 2003).

In the service and maintenance, the work is expected to improve efficiency by accessing databases on-line and reduce human errors by augmenting the real objects with visual information such as annotations, data maps, and virtual models (Navab, 2004). As a solution, an online guided maintenance approach was taken for reducing necessity and dependency on trained workers facing increasingly complex platforms and sophisticated maintenance procedures (Lipson, et al., 1998; Bian, et a., 2006). It has potential to create a new quality of remote maintenance by conditional instructions adjusting automatically to conditions at the maintenance site, according to input information from the machine and updated knowledge at the manufacturer. The service and maintenance of nuclear power plants also require workers to minimize the time for diagnostics (troubleshooting and repair) and comply with safety regulations for inspection of critical subsystems. The context-based statistical pre-fetching component was implemented by using document history as context information (Klinker, et al., 2001). The pre-fetching component records each document request that was made by the application, and stores the identifier of the requested document in a database. The database entries and the time dependencies are analyzed for prediction of documents suitable for the current location and work of the mobile workers.

Early work at Boeing in the assembly process indicated the advantages of HUD technology in assembling cable harnesses (Caudell and Mizell, 1992). Large and complex assemblies are composed of parts, some of which are linked together to form subassemblies. To identify the efficient assembly sequence, engineers evaluate whether the assembly operation is feasible or difficult and edit the assembly plan. An interactive evaluation tool using AR was developed to attempt various sequencing alternatives of the manufacturing design process (Sharma, et al., 1997; Raghavan, et al., 1999). On the other hand, an AR-based furniture assembly tool was introduced for assemblers to be guided step-by-step in a very intuitive and proactive way (Zauner, et al., 2003). The authoring tool was also developed offering flexible and re-configurable instructions. An AR environment allows engineers to design and plan assembly process through manipulating virtual prototypes at the real workplace, which is important to identify the drawbacks and revise the process. However, the revision of the design and planning is time-consuming in the large-scale assembly process. Hierarchical feature-based models were applied updating the related feature models in stead of the entire model. This results in computational simplicity offering a real-time environment (Pang, et al., 2006)

### 4.2 Effects
Compared to conventional printed manuals, HUD-based CAI using an HMD has the following advantages:

1) Hands-free presentation
An HMD presents information with a display mounted on the user's head. Therefore, although cables are attached to supply electric power and data, both hands can freely be used for a task.

2) Reduction of head and eye movements
Superimposing virtual objects onto real-world scenes enables users to view both virtual objects and real scenes with less movement of the eyes and the head. Spatial separation of the display beyond 20 deg. involves progressively larger head movements to access of visual information (Previc, 2000), and information access costs (effort required to access information) increase as spatial separation increases (Wickens, 1992).

3) Viewpoint-based interaction
Information related to the scenes detected by a camera attached to an HMD is presented to the user. Unlike printed media, there is no need for the user to search for the information required for a specific task. The user simply looks at an object, and the pertinent information is presented at the display. This triggering effect enables efficient retrieval of information with little effort by the user (Neumann and Majoros, 1998).

While many systems have been designed based on implicit assumptions that HUDs improve user performance, little direct empirical evidence concerning their effectiveness has been collected. An inspection scenario was examined in three different conditions: an optical see-through AR, a web browser, and a traditional paper-based manual (Jackson, et al., 2001). They found that the condition of the paper manual outperformed those of the others. In a car door assembly, the experimental results showed that the task performance depended on degree of difficulty on the assembly tasks (Wiedenmaier, et al., 2003). The AR condition wearing an HMD was more suitable for the difficult tasks than the paper manual condition, whereas the performance had no significant difference for the easy tasks between the two conditions.

There has been an investigation of how effectively information is accessed in annotated assembly domains. The effectiveness of spatially-registered AR instructions was compared to the other three instructions: a printed manual, CAI on an LCD monitor, and CAI on a see-through HMD, in experiments on a Duplo-block assembly (Tang, et al., 2003). The results showed that the spatially-registered AR instructions improved task performance and relieved mental workload on assembly tasks by overlaying and registering information to the workspace in a spatially meaningful way.

## 5. Case Study

We applied a HUD-based CAI to a support system for the operation of a transportable earth station containing many pieces of equipment used in satellite communications (Tanaka and Kondo, 1999). The transportable earth station was designed so that non-technical staff could manage the equipment in cooperation with a technician at a hub station. However, operating unfamiliar equipment was not easy for them, even though a detailed instruction manual was available. One of the basic problems staff has during the operation was to understand what part of the instruction manual related to which equipment and then figuring out the sequence of steps to carry out a procedure. Another problem was that the transmission at a session leaves little room for error in operating the equipment because mistakes of the transmission operation may give serious damage to the communication devices of the satellite.

### 5.1 Transportable Earth Station

A transportable earth station has been constructed as an extension of the inter-university satellite network that is used for remote lectures, academic meetings, and symposia in higher education, to exchange audio and video signals. The network now links 150 stations at universities and institutes. The transportable earth station has the same functionality as the original stations on campus but can be transported throughout Japan.

Figure 4 (a) shows a photograph of the transportable earth station. The van carries transmitting-receiving devices, video-coding machines, a GPS-based automatic satellite-acquisition system, and various instruments for measurements. The operator has to manage these pieces of equipment with the appropriate procedures and perform the adjustments and settings required for satellite communications. The uplink access test involves the operation of the transmitters and receivers shown in Figure 4 (b), and this requires some specialized expertise and error-free performance.

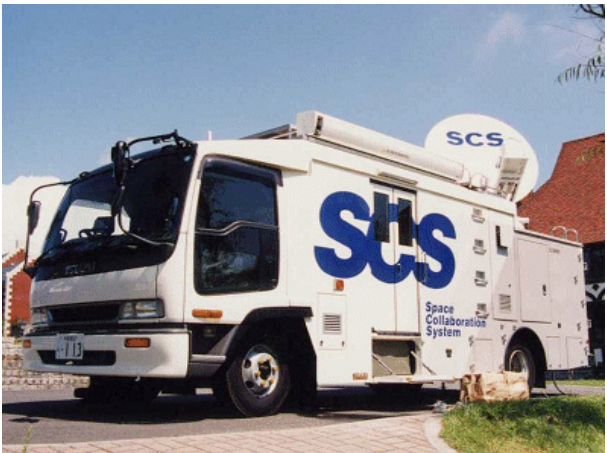
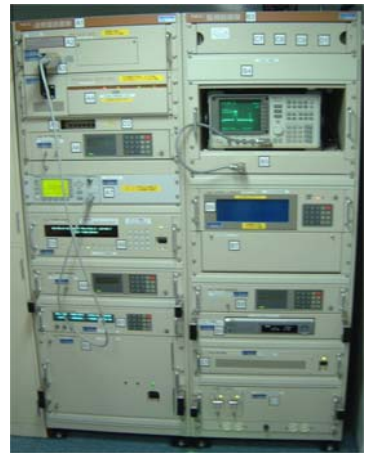

(a) Van　　　　　　　　　　　　　　　　　　　　　　(b) Equipment
Fig. 4. Transportable earth station

### 5.2 HUD-based CAI System

Our HUD-based CAI system was originally designed to improve operation of the transportable earth station. Here, the outline of the prototype system assumes that the system will be used to operate the pieces of equipment in the transportable earth station, though the experiment described in the next section was done under laboratory conditions.

Figure 5 shows a schematic configuration of our prototype system. A compact camera attached to the user's head captures images from the user's viewpoint. These images are transferred to a PC through a DV format. Identification (ID) patterns registered in advance are stuck on the equipment, and each marker is detected in the video images using ARToolkit (Kato, et al., 2000), which is a C/OpenGL-based open-source library that detects and tracks objects using square frames.

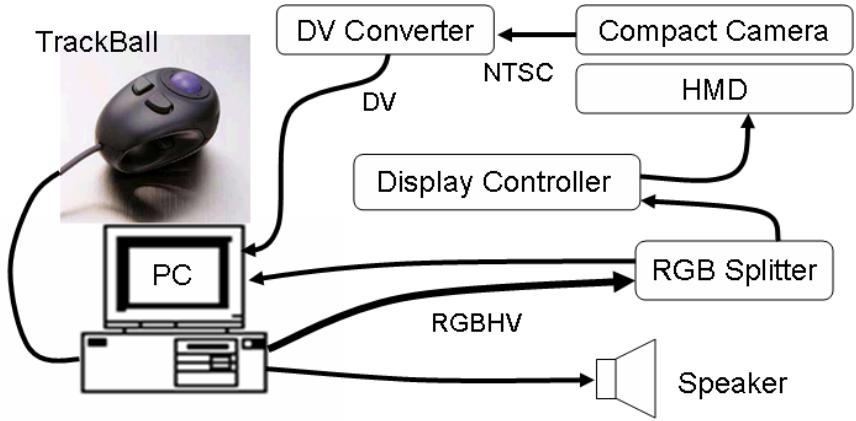

Fig. 5. System configuration

When the marker pattern is found in the registered list, ARToolkit identifies the piece of equipment and the appropriate instructions are presented to the user via an HMD. At the same time, the name of the equipment is stated audibly by a recorded voice to alert the user and make sure he or she works on the right piece of equipment. A square marker centered in or close to the center of the scene is detected, as several markers are present in the same scene. A trackball is used to control pages by, for example, scrolling pages, sticking a page, and turning pages.

**5.3 Software Architecture**
Figure 6 shows the software architecture of the prototype system. The software consists of two parts that have a server-client relationship: image processing and display application. The server and client exchange data using socket communications with UDP/IP. The server-client architecture enables the load to be distributed to two processors, though the prototype system was implemented on one PC.
It is common for graphical signs or simple icons to be presented with spatial registration to real objects in assembly tasks. In our case, however, using such graphical metaphors is insufficient to represent the amount of information because detailed operating instructions for the equipment should be provided based on conditions. Such documents contain too much information to be spatially registered with a single piece of equipment. Unfortunately, the resolution and field of view are currently limited in commercially available HMDs. Therefore, we displayed the manual documents with large fonts, sacrificing spatial registration to the real equipment. The lack of spatial registration may not be a problem for us because, unlike aircraft landing simulation, there is no differential motion in the manipulation of the equipment and the background scene is stable.

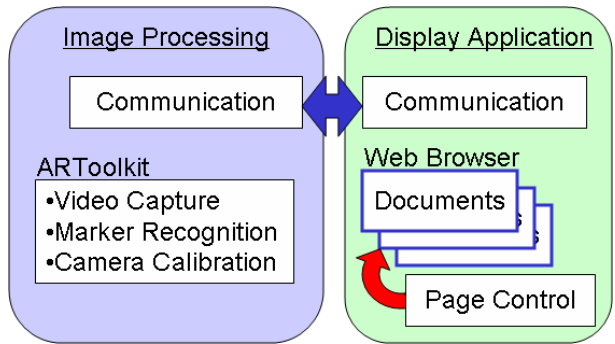

Fig. 6. Software architecture

Pages presented on the HMD are formatted in HTML, which enables a Web browser to be used to display multimedia data and makes instruction writing easier. Writing content for this kind of application, which has usually required programming skills and graphics expertise, is often costly and time consuming. Instructions must be customized to each piece of equipment and sometimes need to be revised, which may greatly increase the workload. The use of a Web browser enables fast implementation and flexible reconfiguration of the component elements in the instructions.

**5.4 Implementation**
The prototype system was implemented on a 2.8-GHz Pentium 4 PC with a 512-MB memory. The video frame rate of the image processing was roughly 30 frames per second. A single-sided HMD (SV-6, produced by Micro Optical) was installed as shown in Fig. 7, attaching a compact camera. The HMD has a viewing angle of roughly 20 degrees in the diagonal and the pinhole camera has a viewing angle of 43 degrees. The HMD, including the camera, weighs 80 g.

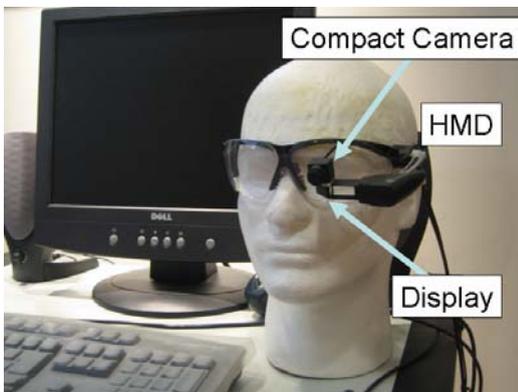
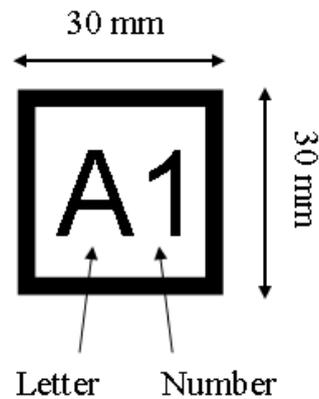

Fig. 7. HMD with a camera          Fig. 8. Sample ID marker

Figure 8 shows a sample ID marker. When the surroundings are too dark, the camera images become less clear, degrading marker detection and recognition accuracy. Although the recognition rate depends on illumination, no recognition error has been observed at a viewing distance within 80 cm when the illumination environment is the same as that in which the marker patterns were registered.

## 6. Experiment

We conducted an experiment that compared the performance of participants using a HUD system, a laptop PC, and a paper manual. We hypothesized that the HUD system would make users receive the HUD profits (hands-free environments, reduction of head and eye movements, and awareness of real objects) and difficulty viewing information on an HMD. We expected that these would affect the time required to perform tasks. Figure 9 shows photos of the experiment being carried out: a) HMD system, b) laptop PC, and c) paper manual, respectively.

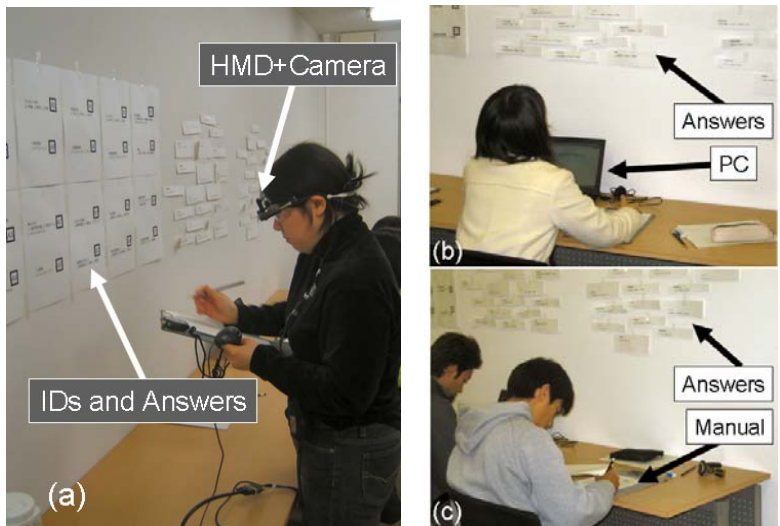

Fig. 9. Experimental tasks (a: HMD system, b: laptop PC, and c: paper manual).

### 6.1 Method
Thirty-five people participated in the experiment. The participants were undergraduate and graduate students who had normal vision and no previous experience of HUD-based CAI.
The task was to read articles and answer questions about the articles. The questions were multiple choices and had three possible answers each. Participants were instructed to read 20 articles with each instruction media. For each article, participants were required to write down the answer and the times when they found the answer possibilities and when they finished answering the questions. The participants were instructed to complete the task as quickly and accurately as possible.

In the experiment, slips of paper with possible answers were taped to the wall in front of the participant. In the HUD system, each possible answer slip has an ID marker, and articles and questions are presented on the HMD. In the PC and the paper manual, articles and questions are presented on the PC monitor and paper sheets, respectively.

All three media used the same format to present the articles, but there were differences among the media in how the questions and answers were presented. In the HUD system and the PC, the title and marker number were displayed in blue at the top of the page, and the headings of the article were displayed in black below the title. These headings were presented in larger fonts on the HMD (24 pt) than on the PC monitor (12 pt). When the participant centers the marker on the display, the article is identified, and the article's top page is presented on the HMD. While reading an article, a participant presses a button of the trackball to hold the page.

The time required to find and complete each article were recorded for all 20 articles. Finding time is defined as the time it took the participant to find the possible answers for a question, and completion time is defined as the time it took the participant to finish answering the question. These times were recorded by the participants themselves using a stopwatch, and participants were asked at a preference test about which media was the best for performing the task. The questions are listed in Table 1.

|   | Question |
|---|----------|
| 1 | The instruction medium was easy to use. |
| 2 | Practice was not necessary. |
| 3 | Looking for possible answers was easy. |
| 4 | Identifying the answer to the question was easy. |
| 5 | The task was not tiring. |
| 6 | The medium was enjoyable. |

Table 1. Preference test questions

The participants were divided into three groups and each group started performing the task using one of the three different instruction media. There were three trials with each group using each medium once. For example, if a participant began with the HUD system, he or she would use the laptop PC in the second trial and the paper manual in the third trial. The preference test was conducted immediately after participants had finished all three trials.

For the HUD system to work properly, the position of the HMD and the direction of the compact camera needed to be calibrated. The calibration took approximately two minutes. The participants practiced with the HUD system, and any questions were answered at that time. The trials started when the participants reported feeling comfortable in using the HUD system.

## 6.2 Results

Figure 10 shows the results of the experiment. The bars indicate average times required by participants to complete trials. The black, shaded, and unshaded bars represent times for the HUD system, the laptop PC, and the paper manual, respectively. The error bar on each bar represents the standard deviation.

Analysis of Variance (ANOVA) was carried out on the task time data. Presentation media significantly affected finding time (F[2,32]=39.6, p<0.01). Post hoc analysis for all possible pairs of presentation media showed that the trial with the HUD system was significantly shorter than those with the others. Presentation media also significantly affected work time (the time required to finish the article after the possible answers had been found) (F[2,32]=22.4, p<0.01). Post hoc analysis showed the trial with the HUD system took significantly longer than those with the other media but no significant difference between the PC system and the paper manual. Presentation media also significantly affected completion time (F[2,32]=6.8, p<0.01). Post hoc analysis showed that the trial with the PC took significantly longer than those with the other media, but there was no difference in completion time between the HUD system and the paper manual.

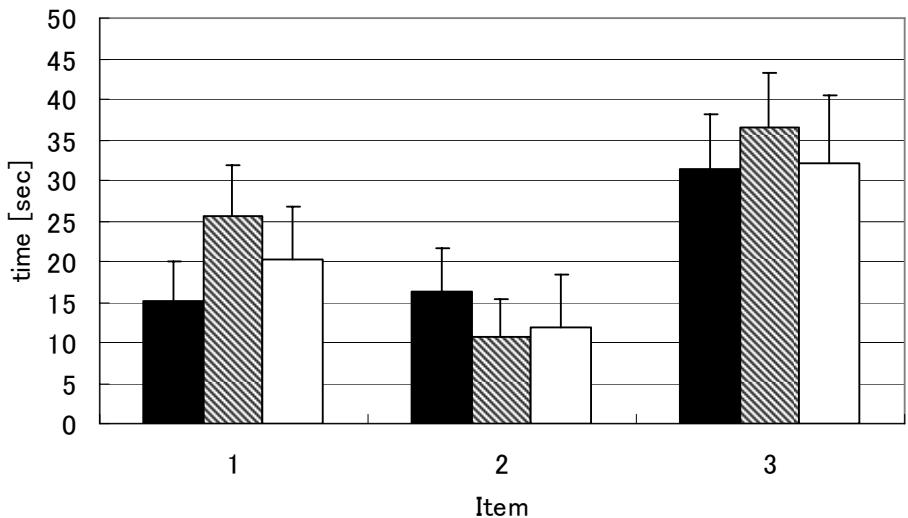

Fig. 10. Experimental results (black: HUD, shaded: laptop PC, unshaded: paper manual)

Figure 11 shows the results of the preference test. Scores for questions by number, as listed in Table 1, are ranged along the horizontal axis. The bars indicate the average score reported by participants. The black, shaded, and unshaded bars represent scores for the HUD system, the laptop PC, and the paper manual, respectively. The error bar on each bar represents the standard deviation.

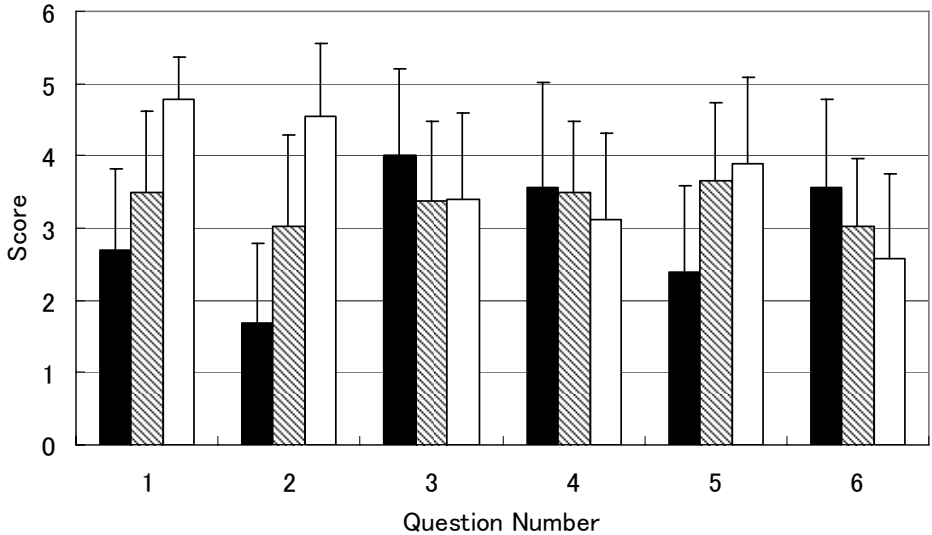

Fig. 11. Preference test results (black: HUD, shaded: laptop PC, unshaded: paper manual)

In answers to questions 1 and 2, related to ease of use and practice, the same tendency was observed: the paper manual was the most preferred instruction medium, the PC was ranked second, and the HUD system was third. In answer to question 5, which asked about how tiring the task was, participants reported that they preferred the PC and the paper manual to the HUD system, but neither of these two was clearly preferred over the other. In answers to question 4, about mental switching, all the scores were comparable. In answers to question 6, participants reported that the HUD system was the most enjoyable, the PC was next, and the paper manual was boring. In answers to question 3, the HUD system was reported to be the most helpful in searching articles, and the other media had comparable scores.

**6.3 Discussion**
Overall, the performance and preference test results did not show that the HUD system was clearly superior to the other media over the whole course of the task. As expected, the participants using the HUD system took less time finding the possible answers and more time reading the articles and the question on the HMD than participants using the other media. The results showed that the HUD system excelled at finding the place of the answer possibilities but seemed to spoil the excellence by careful reading of the articles and questions, which affected the task time. This suggests that the HUD-based CAI system is good at indicating which equipment the user needs to treat but not so suitable for presenting instructional information, because the HMD requires the user to see letters and characters at the limited resolution and field of view.
We also observed that the ID on the answer possibility sheet made it easy to identify the location of the answer on the wall and the article on the HMD. That is, the ID worked as a sign guiding the task procedure.

We found that there was a difference between the experimental results and those obtained in the actual operation of the transportable earth station. In the laboratory experiment, the task completion time was comparable among the three media. In the actual operation, however, people using the HUD-based CAI system performed better than those using the other two media. This was interpreted as a difference of the task or the experimental condition that worked against the HUD system or in favor of the paper manual in this experiment. It was important for the people to check if they were reading the appropriate instructions during the actual operation, because the pieces of equipment were not familiar to them. This situation could work for the HUD system.

## 7. Conclusion

We investigated the role of HUDs in CAI. HUDs have been used in various situations in daily lives by recent downsizing and cost down of the display devices. CAI is one of the promising applications for HUDs. We have developed an HUD-based CAI system for effectively presenting instructions of the equipment in the transportable earth station. This chapter described HUDs in CAI from a viewpoint of human-computer interaction based on the development experience.

First, the basic concept of an HUD was introduced by briefly describing general HUD technology and its relevant applications. An HUD is basically a display medium on which information is presented, allowing a user to simultaneously view a real scene and superimposed information without large movements of the head or eye scans. The HUD has been incorporated into various applications, on which its type depends. We described HUD design types: head-mounted or ground-referenced, optical see-through or video see-through, and single-sided or two-sided types, and discussed their characteristics by comparing each HUD design type.

Second, the features of HUD-based CAI were explained by describing its applications, such as equipment operation, product assembly, and machine maintenance. These HUD-based CAI applications have witnessed the introduction of increasingly complex platforms and sophisticated procedures and are characterized by the real-time presentation of instructional information related to what the user is viewing. Common thought is that HUD-based CAI will increase the efficiency of instructions and reduce errors by properly presenting instructional information based on a user's viewpoint. We discussed the advantages of HUD-based CAI, such as a hands-free environment, reduction of head and eye movements, and awareness of real objects, compared to conventional printed manuals.

Third, a user study with our HUD-based CAI system was reported. Our system provides information using a head-mounted HUD, on which a piece of equipment is identified with identification markers, as the user looks at the piece of equipment that she tries to manipulate. User performance with the system was evaluated during a task in which participants read articles and answered questions about the articles, and this performance was compared to performance with a laptop PC and paper manual. The experimental results for the performance of the HUD-based CAI system showed less time finding pairs of questions and the possible answers and more time selecting one of the possibilities after reading the articles. This suggested that the user would receive the advantages of an HUD, but also have difficulty viewing the information because of the narrow field of view and insufficient resolution in the HUD.

We did not here deal with eye strain and a safety measure. These issues become important when HUDs are used for CAI in actual situations such as a construction site. In general, the display surface and the real objects are different in distance from the user's eyes. This may cause eye strain for a long term use, resulting in visually-induced sickness. Besides, the mechanics of the eyes' protection must work in an accident of bumps between the display device and the periphery, especially for the head-mounted type. These issues have to be addressed before the practical uses of HUD-based CAI.